# Substrate Index Dependence of Energy Levels in $\text{In}_{0.4}\text{Ga}_{0.6}\text{As}/GaAs$ Quantum Dots Applicable to QD-Lasers (a six-band k.p approximation)

Esfandiar Rajaei[*] and Mahdi Ahmadi Borji[**]

Department of Physics, University of Guilan, Rasht, Iran

*Corresponding author, E-mail: Raf404@guilan.ac.ir

**E-mail: Mehdi.p83@gmail.com

## ABSTRACT

Quantum dot lasers have been the focus of researchers due to their interesting optical properties owing to quantum confinement of carriers. In epitaxial quantum dots formed on a substrate, band-edge diagrams sound to more complex than simple bulk materials because of the important role of strain. Strain tensor is strongly dependent to lattice mismatch. Various substrate indexes have been used in the laser devices, and a number of research groups elaborate to find the best substrate index. In this research band edge, energy levels and strain effects of cubic $\text{In}_{0.4}\text{Ga}_{0.6}\text{As} - \text{GaAs}$ quantum dots are studied by 6-band k.p model and their dependence to substrate index is investigated. It is shown that change of substrate index affects the band gap and transition energy of electrons and holes. Moreover, we show that components of strain tensor are strongly sensitive to surface index, and symmetry breaks along space. Our results appear to be in very good consonance with represented formula and similar researches.

Keywords: substrate index, quantum dot laser, band edge, strain tensor.

## I. INTRODUCTION

Quantum Dot (QD) semiconductor lasers have been the focus of many investigations thanks to their optical properties arising from the quantum confinement of electrons and holes [1-4]. By now, QD materials have found very promising applications in optical amplifiers and semiconductor lasers [5-8]. Effects of different parameters, namely, substrate index [9, 10], working temperature [11-14], size [15-17], stoichiometric percentage [18], strain effect [19, 20], … are argued to be important in the performance of QD-Lasers. Thus, a detail and precise study of the effects of these factors can lead to future high performance lasers.

InGaAs/GaAs devices are paid attention in a number of publications due to their interesting attributes [4, 21-25]. In quantum confined hetero-structures with materials of different lattice constants, band edge diagrams show more complexity than bulk materials because of the important role of strain. Strain tensor strongly depends on the elastic properties of connected materials, geometry of the quantum dot, and lattice mismatch [26].

It is shown that optical properties, size distribution, and density and ordering of QDs are associated with substrate-orientation. In former studies, substrate (001) index was mostly used, however, it was found that enhanced photoluminescence intensity with sharper peaks and better QD alignment result from (311) surface of the substrate. These findings agreed with future experimental researches too. Thus, many investigations elaborate to find the best substrate index [10].

In this paper we are going to study the effect of substrate index on properties of $In_{0.4}Ga_{0.6}As$ quantum dots grown on different substrate indexes through $6 \times 6$ k.p technique. The rest of this paper is organized as follows: in section II the model and simulation details are presented. Our results will be discussed in section III in two parts related to band edge (III.A) and strain tensor (III.B); we make the conclusion finally in section IV.

## II. MODEL AND SIMULATION DETAILS

We suppose a cubic $10 \times 10 \times 10 nm^3$ quantum dot on different substrate indexes. Figure 1 shows the profile of a QD and the $0.5 nm$ wetting layer which are surrounded by GaAs. We take indium ratio as 40% as used in many laser applications [27]. Both GaAs and InAs have the zinc-blende structures with direct band gaps. The growth direction was z=[$h$,1,1] perpendicular to $x = [0,1,\bar{1}]$ and $y = [2,\bar{h},\bar{h}]$ [9]. Also for z=[0,0,1] we take $x = [1,0,0]$ and $y = [0,1,0]$. All the simulations in this article are performed under room temperature. We employed Nextnano++ software to implement the 6-band k.p Schrödinger-Poisson solver [26, 28].

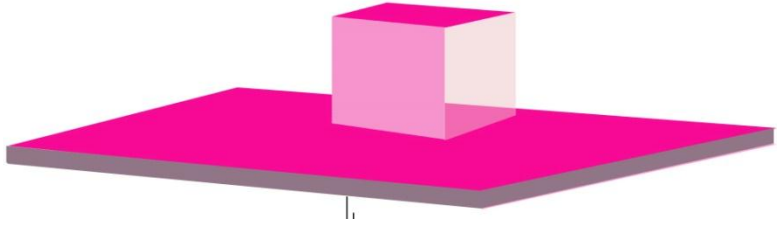

**Figure 1: Profile of a $10 \times 10 \times 10$nm$^3$ cubic In$_{0.4}$Ga$_{0.6}$As QD with a $0.5$nm wetting layer.**

The parameters for bulk materials applied to this article are given in Table 1 [29-31].

**Table 1: Variables applied to our model.**

| Parameter | GaAs | InAs |
|---|---|---|
| Effective electron mass (Γ) | 0.067m$_o$ | 0.026m$_o$ |
| Effective heavy hole mass | 0.5m$_o$ | 0.41m$_o$ |
| lattice constant | 0.565325 nm | 0.60583 nm |
| Nonzero elements of elasticity constants matrix | $C_{11} = 122.1$ $C_{12} = 56.6$ | $C_{11} = 83.29$ $C_{12} = 45.26$ |

|  | $C_{44} = 60$ | $C_{44} = 39.59$ |
|---|---|---|
| Band gap (0K) | 1.424 eV | 0.417 eV |

For $In_{0.4}Ga_{0.6}As$, the effective masses are calculated at room temperature as follows: effective electron mass is obtained $0.04628m_0$ [32], effective heavy-hole mass is $0.47m_0$ [33], and the lattice constant is $a = 0.58153nm$ [34, 35].

## III. RESULTS AND DISCUSSION

### A. Band edge diagram

Figure 2 displays the diagram of conduction and valence band edges for different orientations of the substrate as a function of location. Electronic Γ band edge and Heavy-Hole (HH) edges are seen in the figure associated with their first obtained eigenvalues. Here Δ is defined as the electron-hole transition energy for QD's first eigenvalue. As illustrated, variation of substrate index not only transforms the band edge diagram, but also changes the transition energies. Change in the transition energy, as it is found, is regular; i.e., increase of $h$ in substrate index ($h11$) caused to increased value of transition energy. The ground state recombination energy is applicable to QD lasers. Change of Δ gives rise in a different output wavelength of the laser device. Finally, it is mentioned that band edge of the substrate has also been subjected to change near the interface. GaAs and InAs have direct band-gaps. Band structure is mostly dependent to chemical potential and strongly to strain in heterostructures. In fact, strain is capable of shifting band edge and so affect on the electronic states.

| (a) | (b) | (c) |
|---|---|---|

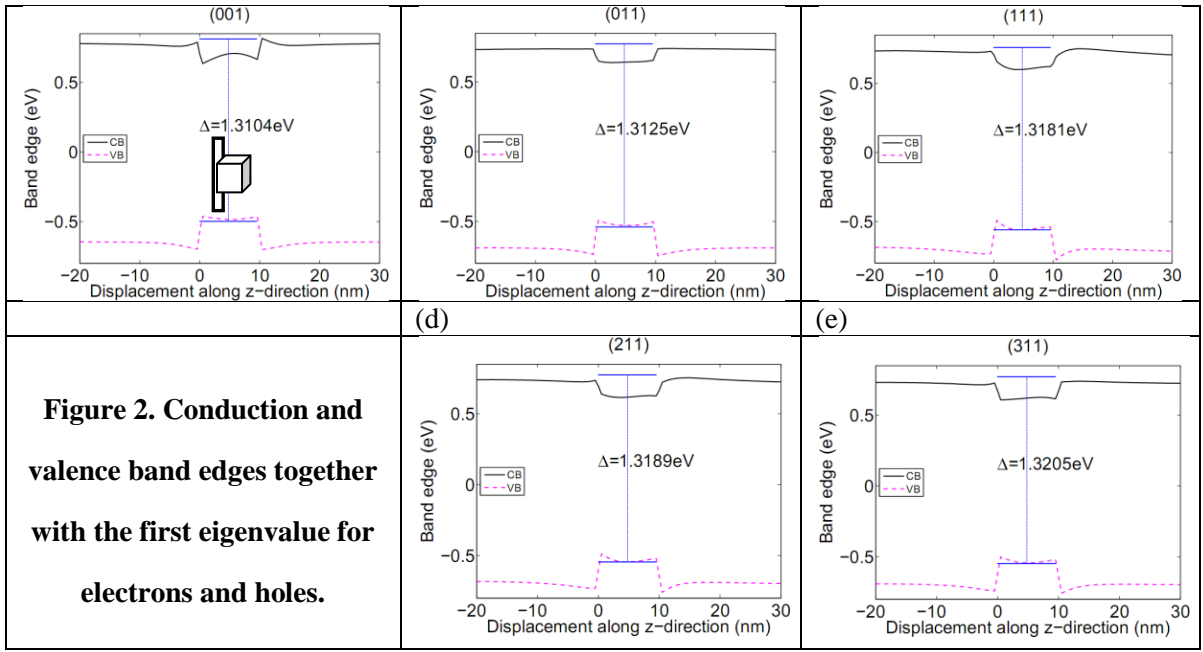

**Figure 2.** Conduction and valence band edges together with the first eigenvalue for electrons and holes.

Better comparison of band edge diagrams for mentioned substrate indexes is shown in figure 3. As it is seen, for conduction and valence band (figures 3a and 3b), surface (111) is associated with the minimum value of band gap and (001) has the maximum value. This shows only a shift in the energy. Moreover, band edge diagram is subjected to change mostly in central points of the quantum dot.

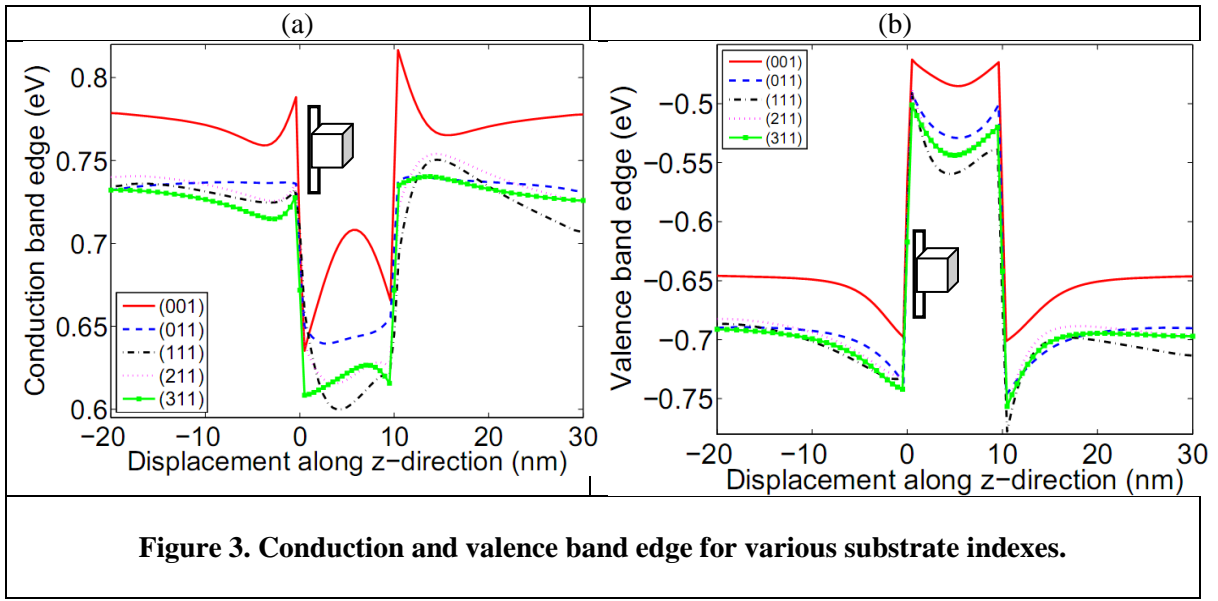

**Figure 3.** Conduction and valence band edge for various substrate indexes.

For a better knowledge of the behavior of band edge and electron-hole states, these parameters are shown in figure 4a for various substrate indexes. As it is seen, the difference is not so remarkable when substrate index changes. However, the lowest band gap belongs to (011) surface and the greatest one is related to (111) (figure 4b). It is also clear that for $h > 0$ rise of $h$ enhances the band gap. In addition, increase of $h$ leads to enhanced electron-hole transition energy.

It has been pointed out in [36] that band gap is dependent to lattice constant. Change of substrate index, as known, changes the lattice constant on the surface on which growth is conducted. They showed that increase of lattice constant results in the decreased ban gap. Also, they show that energy levels in GaAs are strongly strain-dependent.

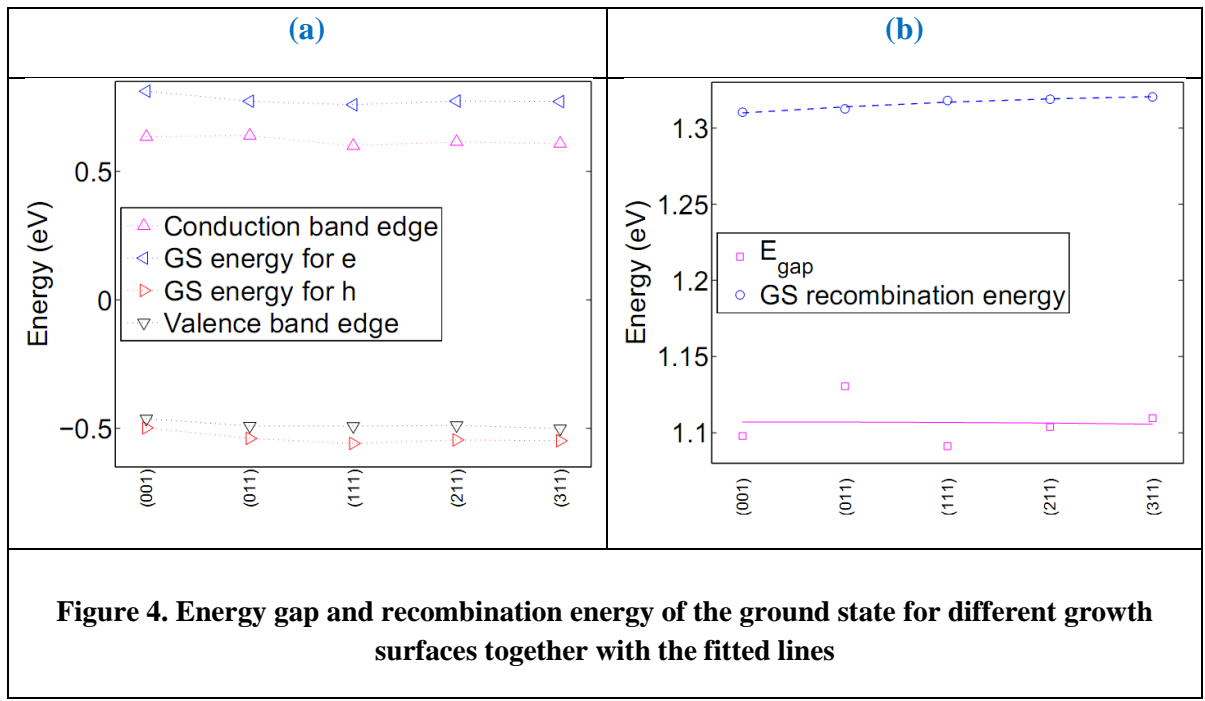

**Figure 4. Energy gap and recombination energy of the ground state for different growth surfaces together with the fitted lines**

**B.  Strain tensor**

Strain plays an important role in optimization of optical and electronic properties of heterostructures. Taking into account length variation in all directions, the strain tensor is as follows:

$$\varepsilon_{ij}(\mathbf{r}) = \frac{\frac{du_i(r)}{dr_j} + \frac{du_j(r)}{dr_i}}{2} \quad (1)$$

where $i, j$ take values 1, 2, 3 which respectively refer to x, y, and z. $r_i$ is the length in direction $i$, and $du_j$ is displacement in direction $j$ due to lattice deformation [9]. Although the distortion matrix $u$ may be non-symmetric, this tensor is real and symmetric (i.e., $\varepsilon_{ji} = \varepsilon_{ij}$). Diagonal components are associated with expansion per unit length along an axis which refers to compressive (tensile) strain if negative (positive); in this situation, angles of the volume element are fixed but the lengths change. Off-diagonal components are related to rotation and shear deformations which lead to variation of angles while the volume remains fixed. The resulting general strain tensor is a $3 \times 3$ matrix as follows:

$$\varepsilon = \begin{bmatrix} \varepsilon_{xx} & \varepsilon_{xy} & \varepsilon_{xz} \\ \varepsilon_{xy} & \varepsilon_{yy} & \varepsilon_{yz} \\ \varepsilon_{xz} & \varepsilon_{yz} & \varepsilon_{zz} \end{bmatrix} \qquad (2)$$

Strain is sensitive to material elastic constants. Strain tensor components are represented in figure 5 for the same previous substrate indexes. Obviously, substrate index remarkably changes the strain components. It can be noticed that the mismatch due to the change in substrate lattice constant while changing the growth surface has lead to jump in the strain tensor at interfaces. Also, in all the figures, as expected, $\varepsilon_{xx}$ coincides with $\varepsilon_{yy}$. This result seems logical, since strain is dependent to lattice constant which is itself dependent to the surface on which growth has been conducted.

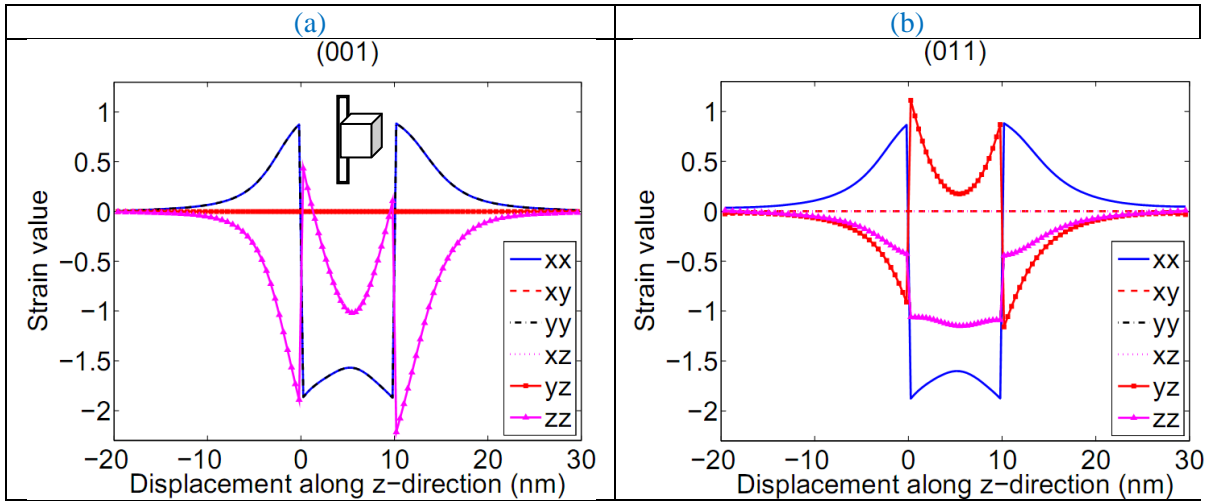

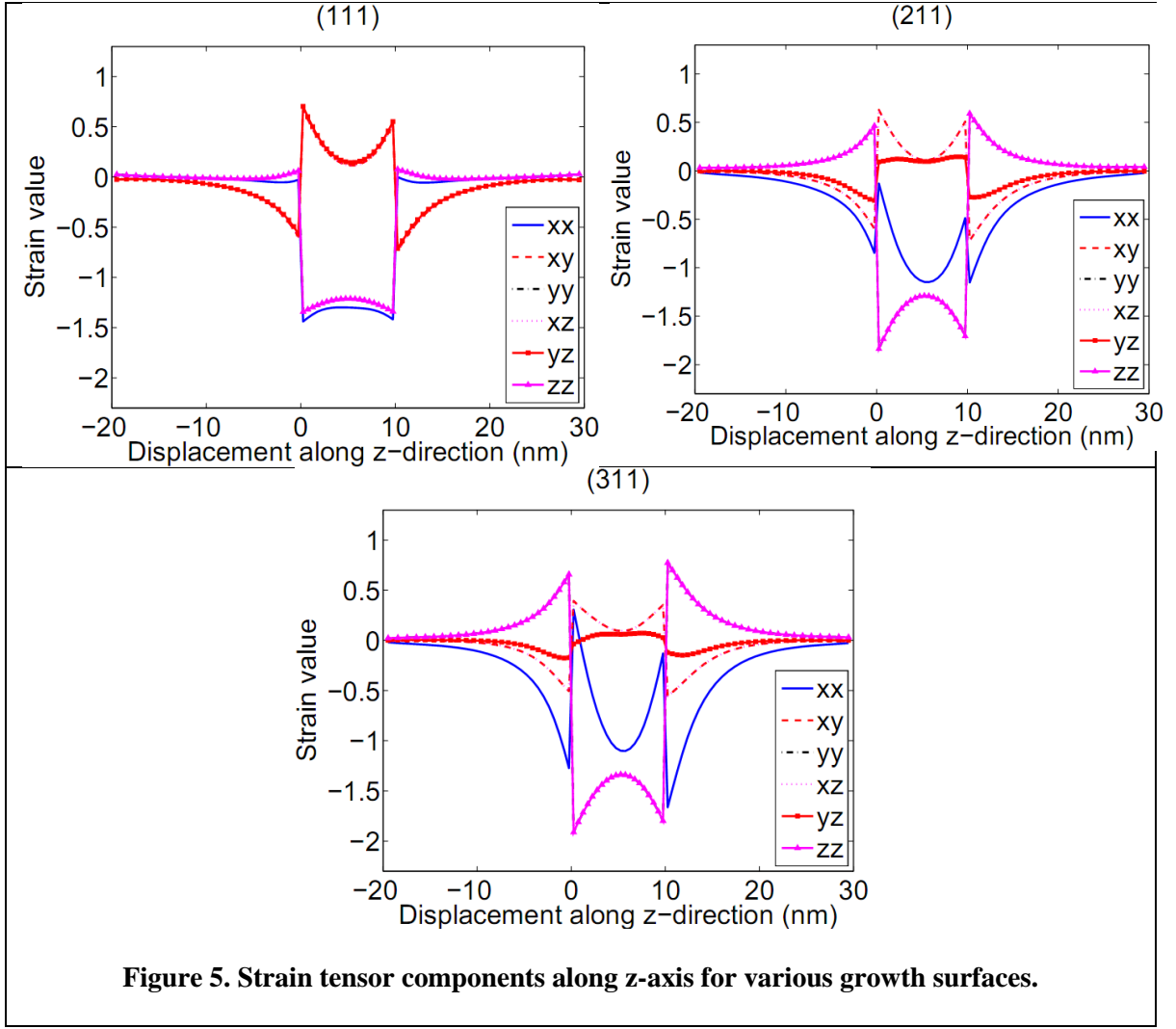

**Figure 5. Strain tensor components along z-axis for various growth surfaces.**

The uniaxial strain component $\varepsilon_{zz}$ perpendicular to the interface and for substrate index (001) is obtained as:

$$\varepsilon_{zz} = \varepsilon_\perp = -\frac{2C_{12}}{C_{11}}\varepsilon_\parallel \qquad (3\text{-a})$$

which is obtained by vanishing $\sigma_{zz}$. Also, biaxial strain parallel to interface reads [37]:

$$\varepsilon_{xx} = \varepsilon_{yy} = \varepsilon_\parallel = \frac{a_{substrate} - a_{QD}}{a_{QD}} \qquad (3\text{-b})$$

in which $a_{substrate}$ and $a_{QD}$ are respectively the substrate and quantum dot lattice constants. $C_{ij}$ are elements of the matrix of elasticity constants which correlates stress $\sigma$ to strain by Hooke's law (i.e., $\sigma = C\varepsilon$) [36, 38]. This relation for our cubic zinc-blende crystal is given as:

$$\begin{bmatrix} \sigma_{xx} \\ \sigma_{yy} \\ \sigma_{zz} \\ \sigma_{xy} \\ \sigma_{xz} \\ \sigma_{yz} \end{bmatrix} = \begin{bmatrix} C_{11} & C_{12} & C_{12} & 0 & 0 & 0 \\ C_{12} & C_{11} & C_{12} & 0 & 0 & 0 \\ C_{12} & C_{12} & C_{11} & 0 & 0 & 0 \\ 0 & 0 & 0 & 2C_{44} & 0 & 0 \\ 0 & 0 & 0 & 0 & 2C_{44} & 0 \\ 0 & 0 & 0 & 0 & 0 & 2C_{44} \end{bmatrix} \begin{bmatrix} \varepsilon_{xx} \\ \varepsilon_{yy} \\ \varepsilon_{zz} \\ \varepsilon_{xy} \\ \varepsilon_{xz} \\ \varepsilon_{yz} \end{bmatrix} \quad (4)$$

Also, in general, when substrate index changes, strain tensor elements are obtained by [39]

$$\varepsilon_{\parallel} = \varepsilon_{xx} = \varepsilon_{yy} = \frac{(a_{substrate} - a_{QD})}{a_{QD}} \quad \text{(5-a)}$$

$$\varepsilon_{xy} = \varepsilon_{yx} = 0 \quad \text{(5-b)}$$

$$\varepsilon_{xz} = \varepsilon_{zx} = \frac{d_4 d_8 - d_6 d_7}{d_4 d_5 - d_6^2} \varepsilon_{\parallel} \quad \text{(5-c)}$$

$$\varepsilon_{yz} = \varepsilon_{zy} = \frac{d_7 - \frac{d_6 \varepsilon_{zx}}{\varepsilon_{xx}}}{d_4} \varepsilon_{\parallel} \quad \text{(5-d)}$$

$$\varepsilon_{\perp} = \varepsilon_{zz} = \frac{d_1 - \frac{2C_{34}\varepsilon_{yz}}{\varepsilon_{xx}} - \frac{2C_{35}\varepsilon_{zx}}{\varepsilon_{xx}}}{C_{33}} \varepsilon_{\parallel} \quad \text{(5-e)}$$

Where growth direction is along $z$ axis and x, y are parallel to the substrate. Here,

$$d_1 = -C_{13} - C_{23} \quad \text{(6-a)}$$

$$d_2 = -C_{14} - C_{24} \quad \text{(6-b)}$$

$$d_3 = -C_{15} - C_{25} \quad \text{(6-c)}$$

$$d_4 = 2C_{33}C_{44} \tag{6-d}$$

$$d_5 = 2C_{33}C_{55} \tag{6-e}$$

$$d_6 = 2C_{33}C_{45} - C_{34}C_{35} \tag{6-f}$$

$$d_7 = C_{33}d_2 - C_{34}d_1 \tag{6-g}$$

$$d_8 = C_{33}d_3 - C_{35}d_1 \tag{6-h}$$

Figure 6 depicts the same strain tensor components for different substrate indexes but along x-axis. For indium percentage of 40, it is expected for the mismatch to be 2.8665%. Therefore, strain maximum value cannot be greater than this value. As it is seen, the maximum value corresponds to that of (001) for $\varepsilon_{zz}$. The other point is that for all $(h11)$ surfaces, $\varepsilon_{yy}$ and $\varepsilon_{zz}$ change and decease of one is synchronized with an increase in the other one. Also, their peak decreases by increasing $h$-value. This reverse relationship for trends is true for $\varepsilon_{xy}$ and $\varepsilon_{xz}$ too. i.e., when one increased the other one drops-down and vice versa. For these two components, the peak increases by $h$ increase.

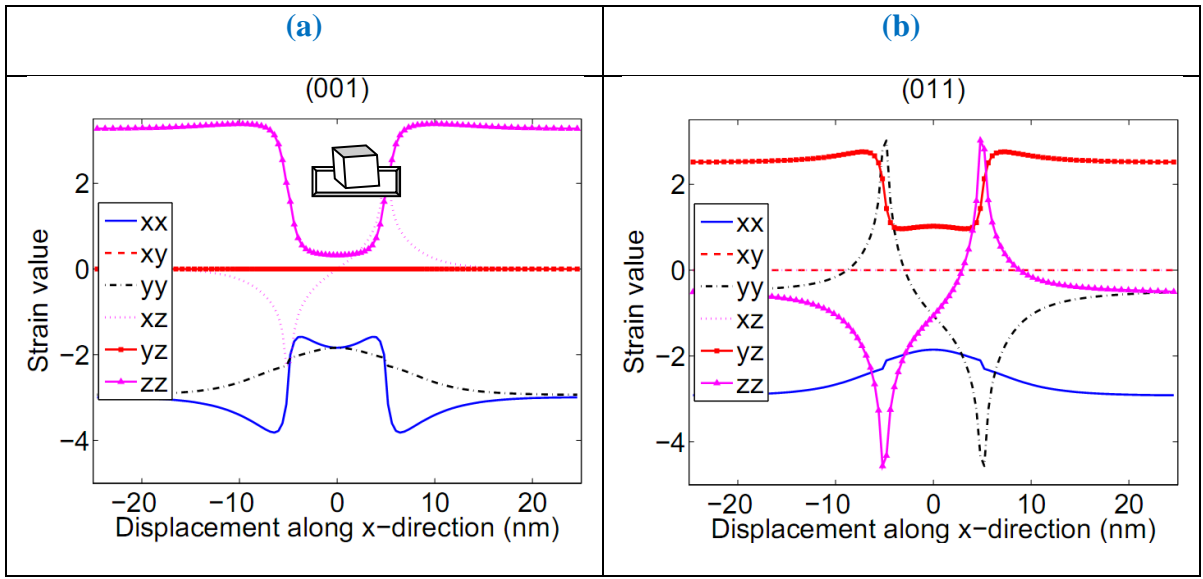

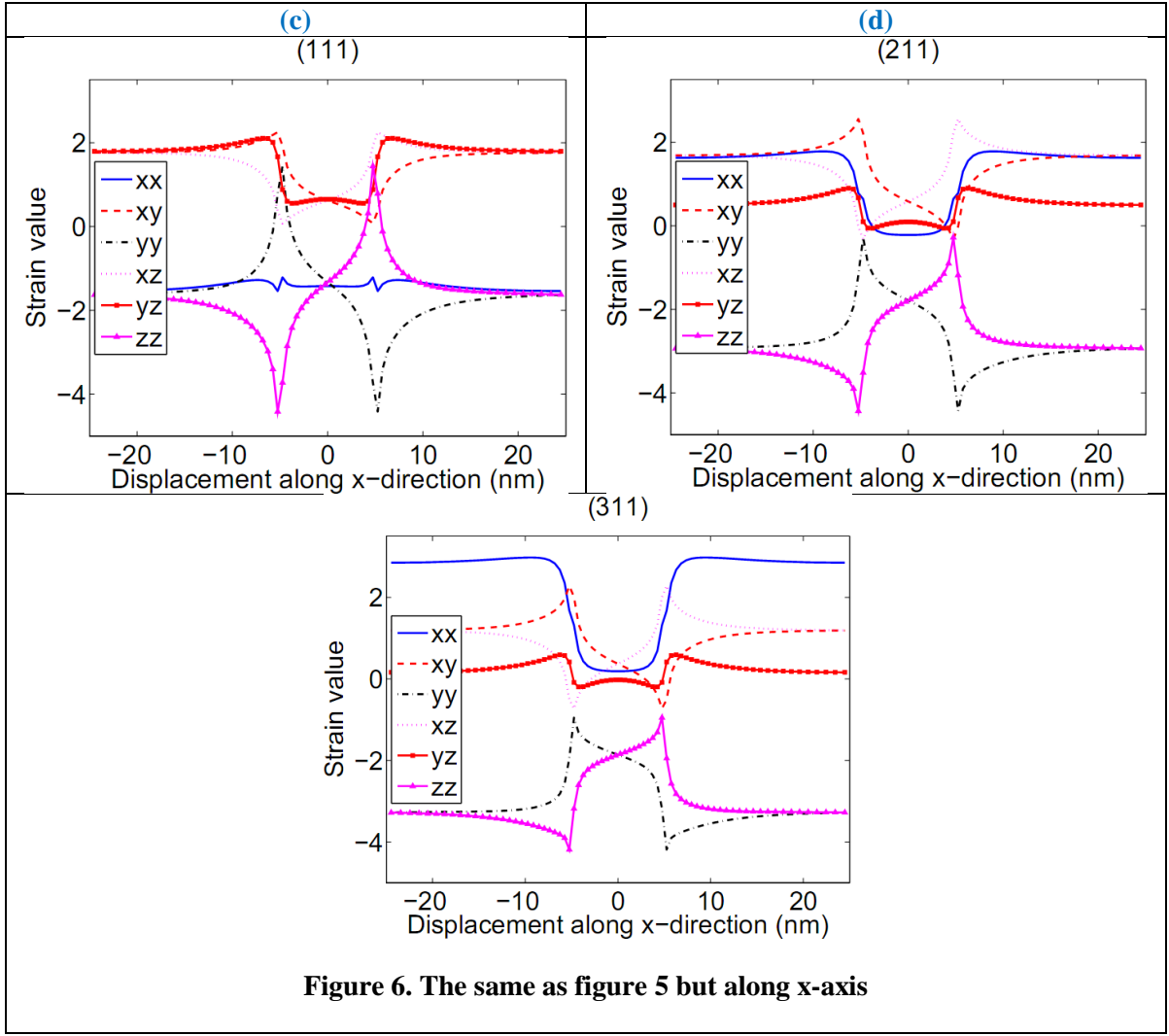

**Figure 6. The same as figure 5 but along x-axis**

## IV. CONCLUSION

We conducted a theoretical study of the effect of substrate orientation on band edge, energy levels and strain effects of cubic $In_{0.4}Ga_{0.6}As - GaAs$ quantum dots. It was shown that change of substrate index affects on the band gap and electron-hole transition energy. Moreover, components of strain tensor were found to be significantly dependent to surface index and while our device was a symmetric system, the symmetry broke down. Our results appear to be in very good consonance with similar researches.

# Acknowledgement

We thank Prof. S. Farjami Shayesteh for helpful comments about this work.